\documentclass[nofootinbib,showpacs,prd,superscriptaddress]{revtex4}

\usepackage{amsmath}
\usepackage{amsfonts}
\usepackage{amssymb}
\usepackage{graphicx}
\usepackage{color}
\def\be{\begin{equation}}
\def\ee{\end{equation}}
\def\bea{\begin{eqnarray}}
\def\eea{\end{eqnarray}}

\begin{document}

\title{Characterizing repulsive gravity with curvature eigenvalues }

\author{Orlando Luongo}
\affiliation{Dipartimento di Fisica, Universit\`a di Napoli ''Federico II'', Via Cinthia, Napoli, Italy.}
\affiliation{Istituto Nazionale di Fisica Nucleare (INFN), Sez. di Napoli, Via Cinthia, Napoli, Italy.}
\affiliation{Instituto de Ciencias Nucleares, Universidad Nacional Autonoma de M\'exico (UNAM), Mexico.}

\author{Hernando Quevedo}
\affiliation{Instituto de Ciencias Nucleares, Universidad Nacional Autonoma de M\'exico (UNAM), Mexico.}
\affiliation{Dipartimento  di Fisica and Icra, ``Sapienza" Universit\`a di Roma, Piazzale Aldo Moro 5, I-00185, Roma, Italy.}

\begin{abstract}
Repulsive gravity has been investigated in several scenarios near compact objects by using different intuitive approaches. Here, we propose an invariant method to characterize regions of repulsive gravity, associated to black holes and naked singularities. Our method is based upon the behavior of the curvature tensor eigenvalues, and leads to an invariant definition of a \emph{repulsion radius}. The repulsion radius determines a physical region, which can be interpreted as a repulsion sphere, where the effects due to repulsive gravity naturally arise. Further, we show that the use of effective masses to characterize repulsion regions can lead to coordinate-dependent results whereas, in our approach, repulsion emerges as a consequence of the spacetime geometry in a completely invariant way.  Our definition is tested in the spacetime of an electrically charged Kerr naked singularity and in all its limiting cases. We show that a positive mass can generate repulsive gravity if it is equipped with an electric charge or an angular momentum. We obtain reasonable results for the spacetime regions contained inside the repulsion sphere whose size and shape depend on the value of the mass, charge and angular momentum. Consequently, we define repulsive gravity as a classical relativistic effect by using the geometry of spacetime only.
\end{abstract}


%\date{\today}

\maketitle

%%%%%%%%%%%%%%%%%%%%%%%%%%%%%%%%%%%%%%%%%%%%%%%%%%%%%%%%%%%%%%%
%%%%%%%%%%%%%%%%%%%%%%%%%%%%%%%%%%%%%%%%%%%%%%%%%%%%%%%%%%%%%%%
\section{Introduction}

In Einstein's general relativity, naked singularities have been shown to exist under quite general assumptions as exact solutions of the corresponding field equations \cite{primaref}. Indeed, each one of the black hole solutions possesses a naked singularity counterpart that appears as soon as the black hole parameters violate the condition for the existence of an event horizon. Many other naked singularity solutions are known for which no black hole counterpart exists \cite{solutions}. This means that naked singularities are well-defined mathematical solutions of Einstein's equations. The interesting question is whether these mathematical solutions describe physical configurations that could exist in Nature. The answer to this question is negative, if the cosmic censorship conjecture \cite{penrose} turns out to be true. Indeed, according to this conjecture,
a physically realistic gravitational collapse, which evolves from a regular initial state,
can never lead to the formation of a naked singularity; that is, all singularities resulting from a gravitational collapse should always be enclosed within an event horizon so that they cannot be observed from outside the horizon. Many attempts have been made
to prove the cosmic censorship hypothesis with the same mathematical rigor used to show the inevitability of
singularities in general relativity \cite{hawking}. So far, no general proof has been formulated and the investigation of several particular scenarios of gravitational collapse corroborate the correctness of the conjecture. Other studies studies \cite{naked}, however,
indicate that under certain circumstances naked singularities can appear during the evolution of a mass distribution into a gravitational collapse. It has been shown that in an inhomogeneous collapse,
there exists a critical degree of inhomogeneity below which black holes
form. Naked singularities appear if the degree of inhomogeneity is bigger than the critical value \cite{citazione1}. The collapse speed and the shape of the collapsing object are also factors playing an
important role in the determination of the final state.  Naked singularities form
more frequently if the collapse occurs very rapidly and the object is not exactly spherically
symmetric. These results indicate that the probability of existence of naked singularities cannot be neglected {\it a priori}. In view of this fact, one of the main goals of our work is to investigate the physical effects around naked singularities, showing possible regions of repulsive gravity. The simplest case of naked singularity is the Schwarzschild spacetime with negative mass. The naked singularity is situated at the origin of coordinates and the entire spacetime represents a repulsive gravitational field. One can interpret this case as represented by an effective mass which   follows from the corresponding Newtonian limit,  and is always negative. This interpretation can be generalized to other spacetimes, leading however to certain difficulties as we will show in Sec. \ref{sec:meff}. Other definitions of effective mass show a completely different behavior \cite{def84}.  One may also try to define repulsive gravity in terms of invariants of the curvature tensor \cite{def89}, although several problems appear. In particular, the Schwarzschild naked singularity is characterized by the same curvature invariants as the Schwarzschild black hole. Analogously, the use of null cones to define repulsive gravity is also not free of difficulties \cite{def08}.

The main purpose of the present work is to propose and test an approach based upon an invariant representation of the curvature tensor and its eigenvalues. A preliminary study of this idea was presented in \cite{lq12}. The region where repulsive gravity can become dominant is defined in an invariant way by considering the behavior of the curvature tensor eigenvalues. To this end, the extremal points of the eigenvalues are considered as indicating a change in the behavior of gravity.  We will show that this definition can be applied to different types of naked singularities, and in all the analyzed cases the obtained results are physically reasonable. In fact, we will see that the repulsion region is always located at a very short distance from the central gravity source.

The paper is organized as follows. In Sec. \ref{sec:meff}, we investigate the behavior of the effective mass as a possibility to determine the regions of spacetime where repulsive gravity can occur. In Sec. \ref{sec:cur}, we introduce the formalism to find the eigenvalues of the curvature tensor in an invariant manner, and define a repulsion radius that determines the region where repulsive gravity can become dominant. In Sec. \ref{sec:bhs}, we test our definition of repulsive gravity in the case of naked singularities with  black hole counterparts.
Finally, in Sec. \ref{sec:con} we discuss our results.

%%%%%%%%%%%%%%%%%%%%%%%%%%%%%%%%%%%%%%%%%%%%%%%%%%%%%%%
%%%%%%%%%%%%%%%%%%%%%%%%%%%%%%%%%%%%%%%%%%%%%%%%%%%%%%%
\section{The effective mass}
\label{sec:meff}

The most general black hole spacetime in Einstein-Maxwell theory is described by the Kerr-Newman metric that in Boyer-Lindquist coordinates can be written as \cite{mtw}
\begin{eqnarray}
ds^2 & = &  {\frac{r^2 - 2Mr + a^2+Q^2} {r^2 + a^2\cos^2 \theta}}
(dt - a\sin^2\theta d\varphi)^2 - {\frac{\sin^2 \theta} {r^2 + a^2\cos^2 \theta}}[(r^2 + a^2) d\varphi - adt]^2
\nonumber \\ &&
  - {\frac{r^2 + a^2\cos^2 \theta} {r^2 - 2Mr + a^2+Q^2}}dr^2 -
(r^2 + a^2\cos^2\theta)d\theta^2 \ ,
\label{kn}
\nonumber
\end{eqnarray}
where $M$ is the mass of the rotating central object, $a=J/M$ is the specific angular momentum, and $Q$ represents the total electric charge.
The corresponding electromagnetic vector potential
\be
A=-\frac{Q r}{\Sigma}\left[dt-a\sin^2\theta d\varphi\right]\,,
\ee
indicates that the magnetic field is generated by the rotation of the charge distribution.

The limiting cases of the Kerr-Newman metric are the Kerr metric for $Q=0$, the Schwarzschild metric which is recovered for $a=Q=0$, the Reissner-Nordstr\"om  spacetime for $a=0$, and the Minkowski metric of special relativity for $a=Q=M=0$. The Kerr-Newman  spacetime is asymptotically flat and free of curvature singularities outside a region situated very close to the origin of coordinates. Indeed, the central ring singularity is determined by the roots of the equation
\be
r^2 + a^2 \cos^2\theta =0 \ ,
\ee
and is covered from the outside spacetime by two horizons situated at
\be
r_{\pm} \equiv M \pm \sqrt{M^2 - a^ 2 - Q^2},
\ee
which are real quantities only if the condition $M^2\geq Q^2 + a^2$ is satisfied. In this case, $r_+$ and $r_-$ represent the radius of the outer and
inner horizon, respectively, and the Kerr-Newman solution is interpreted as describing the exterior field of a rotating charged black hole. In the case $M^2<a^2+ Q^2$,  no horizons exist and the gravitational field corresponds to that of a naked ring singularity.

The Newtonian potential $\Phi$ and the effective mass are usually computed from the metric component $g_{tt}$ as
\be
\lim_{r\to\infty} g_{tt} = 1 - 2 \Phi = 1 - \frac{2M_{eff}}{r} \ .
\ee
In the case of the Kerr-Newman spacetime, we obtain \cite{pap66}
\be
M_{eff} = M - \frac{Q^2}{2r} -\frac{a^2M \cos^2\theta}{r^2}\ ,
\ee
where we neglected all terms proportional to $1/r^3$ and higher. Since the effective mass depends on the radius $r$, it then follows that near the origin of coordinates the effective mass can become negative. The effective mass can be therefore interpreted as a source for generating possible repulsive gravitational fields \cite{lab1,lab2,lab3}.

In the regions where the effective mass is positive, the gravitational field becomes attractive so that the limit of the repulsion region is determined by the zero of the effective mass function, $M_{eff}=0$, i.e.,
\be
r_{rep}^N = \frac{1}{4M}\left(Q^2 + \sqrt{Q^4 + 16 M^2 a^2 \cos^2\theta}\right)\,.
\ee
In the limiting case of the Reissner-Nordstr\"om spacetime ($a=0$), the repulsion region lies inside the sphere with radius $\frac{Q^2}{2M}$. Interestingly, this radius is located inside the classical radius of a charged particle with mass $M$ and charge $Q$,  $r_{class} = \frac{Q^2}{M}$ which is usually interpreted as the radius of a sphere inside which quantum effects become important \cite{lab4}. This fact makes difficult the physical interpretation of the repulsion region as a classical (non quantum) effect. In the case of a Kerr spacetime $(Q=0)$, the repulsion region depends on the azimuthal angle, reaching its maximum value on the axis and its minimum value  on the equatorial plane. Although the above definition of the repulsive gravity region is very simple and intuitive, it is based on assuming the Newtonian potential. Thus, the regions of repulsive gravity would depend on the choice of coordinates, and cannot be defined in a coordinate independent way.

Indeed, some time ago Ehlers \cite{ehlers1,ehlers2} noticed that the computation of the Newtonian potential leads to contradictory results when different coordinate systems are used. Ehlers introduced a general theory (Rahmentheorie) in which Newtonian gravity and Einstein's general relativity are considered as particular theories and the Newtonian limit is mathematically well defined. It can be shown that using Ehlers' Rahmentheorie, the Newtonian limit of the Kerr-Newman metric does not depend on the specific angular momentum \cite{quev90}. In some sense, this result is expected because in Newtonian's gravity the rotation is not a source of gravity. We conclude that the Newtonian approximation is not suitable for investigating repulsive gravity in an invariant manner.

%%%%%%%%%%%%%%%%%%%%%%%%%%%%%%%%%%%%%%%%%%%%%%%%%%%%%%%%%%%%%%%%
%%%%%%%%%%%%%%%%%%%%%%%%%%%%%%%%%%%%%%%%%%%%%%%%%%%%%%%%%%%%%%%%
\section{Curvature invariants and eigenvalues}
\label{sec:cur}

In general relativity, it is necessary to use invariant quantities, e.g. scalars, to describe physical phenomena independently of the choice of coordinates or observers. Since curvature is interpreted as a measure of the gravitational interaction, it is reasonable to use the curvature invariants to analyze the behavior of gravity. From the components of the Riemann curvature tensor, one can build in general 14 functionally independent scalars. In the case of empty space, only four scalars are linearly independent \cite{deb56}. In the case of scalars that are quadratic in the components of the curvature tensor, one can form the following combinations \cite{cbcr02}

\begin{subequations}
\label{aiuto1}
\begin{align}
K_1 = R_{\alpha\beta\gamma\delta}R^{\alpha\beta\gamma\delta}\,,\\
K_2 = [{}^*R{}]_{\alpha\beta\gamma\delta}R^{\alpha\beta\gamma\delta}\,,\\
K_3 = [{}^*R{}^*]{}_{\alpha\beta\gamma\delta}R^{\alpha\beta\gamma\delta}\,,
\end{align}
\end{subequations}

where the asterisk represents dual conjugation. They are known as the Kretschmann invariant, the
Chern-Pontryagin invariant, and the Euler invariant, respectively. Those invariants have been employed also for describing the effects of dark matter and dark energy in late time cosmology \cite{ego,noego}. It has been argued that both the effects of dark matter and dark energy may be described in terms of a single scheme, by means of Eqs. (\ref{aiuto1}). Indeed, for describing repulsive effects due to the antigravitational behavior of dark energy, one may include the Kretschmann invariants as a source for the whole energy momentum tensor.

The behavior of the above invariants has been proposed to determine the ``repulsive domains" of gravity and negative effective masses in curved spacetimes \cite{def89}; however, their quadratic structure does not allow us to consider all possible cases of naked singularities. Indeed, for the Schwarzschild spacetime one obtains $K_1=48M^2/r^6$ whereas $K_2$ and $K_3$ are proportional to $K_1$. Since the change $M\rightarrow -M$ does not affect the behavior of $K_1$, these invariants do not acknowledge the existence of a Schwarzschild naked singularity. Rephrasing it differently, a Schwarzschild black hole and a corresponding naked singularity have the same quadratic invariants. Similar difficulties appear in more general cases like the Kerr and Kerr-Newman naked singularities. It is worth noticing that the Ricci scalar, which is the only linear invariant in the curvature tensor components, is not suitable for investigating the problem of repulsive gravity, because it vanishes identically for all the above naked singularity solutions.

Hence, the need for addressing the problem of using first order invariants to characterize repulsive effects becomes essential in order to describe naked singularities. To this end, here we propose an alternative approach in which the eigenvalues of the curvature tensor play the most important role. There are different ways to determine these eigenvalues \cite{solutions}. Our strategy is to use local tetrads and differential forms. From the physical point of view, a local orthonormal tetrad is the simplest and most natural choice for an observer in order to perform local measurements of time, space, and gravity. Moreover, once a local orthonormal tetrad is chosen, all the quantities related to this frame are invariant with respect to coordinate transformations. The only freedom remaining in the choice of this local frame is a Lorentz transformation. So, let us choose the orthonormal tetrad as
\be
ds^2 = g_{\mu\nu} dx^\mu dx^\nu= \eta_{ab}\vartheta^a\otimes\vartheta^b\ ,
\ee
with $\eta_{ab}={\rm diag}(+1,-1,-1,-1)$, and $\vartheta^a = e^a_{\ \mu}dx^\mu$. The first
\be
d\vartheta^a = - \omega^a_{\ b }\wedge d\vartheta^b\ ,
\ee
and second Cartan equations
\be
\Omega^a_{\ b} = d\omega^a_{\ b} + \omega^a_{ \ c} \wedge \omega^c_{\ b} = \frac{1}{2} R^a_{\ bcd} \vartheta^c\wedge\vartheta^d
\ee
allow us to compute the components of the Riemann curvature tensor in the local orthonormal frame. It is convenient to decompose the curvature tensor in terms of its irreducible parts with respect to the Lorentz group which are the Weyl tensor
\cite{deb56}
\be
W_{abcd}=R_{abcd}+2\eta_{[a|[c}R_{d]|b]}+\frac{1}{6} R\eta_{a[d}\eta_{c]b}\ ,
\label{weyl}
\ee
the trace-free Ricci tensor
\be
 E_{abcd} = 2\eta_{[b|[c}R_{d]|a]} - \frac{1}{2}R\eta_{a[d}\eta_{c]b}\ ,
\label{trace}
\ee
and the curvature scalar
\be
S_{abcd} = -\frac{1}{6} R\eta_{a[d}\eta_{c]b}\ ,
\label{scal}
\ee
where the Ricci tensor is defined as $R_{ab}=\eta^{cd}R_{cabd}$. It is possible to represent the curvature tensor as a  (6$\times$6)-matrix by introducing the bivector indices $A,B,...$ which encode the information of two different tetrad indices, i.e., $ab\rightarrow A$. We follow the convention proposed in \cite{mtw} which establishes the following correspondence between tetrad and bivector indices
\be
01\rightarrow 1\ ,\quad 02\rightarrow 2\ ,\quad 03\rightarrow 3\ ,\quad 23\rightarrow 4\ ,\quad 31\rightarrow 5\ ,\quad 12\rightarrow 6\ .
\ee
This correspondence can be applied to all the irreducible components of the Riemann tensor given in Eqs.(\ref{weyl})--(\ref{scal}) so that the bivector representation can be expressed as
\be
R_{AB} = W_{AB} + E_{AB} + S_{AB}\ ,
\ee
with
\begin{equation}
W_{AB}=\left(
         \begin{array}{cc}
           M & N \\
           N & -M \\
         \end{array}
       \right), \
       E_{AB}=\left(
         \begin{array}{cc}
           P & Q \\
           Q & -P \\
         \end{array}
       \right),\
       S_{AB}=-\frac{R}{12}\left(
         \begin{array}{cc}
           I_{3} & 0 \\
           0 & -I_{3} \\
         \end{array}
       \right).
\label{so31}
\end{equation}

Here $M$, $N$ and $P$ are $(3\times 3)$ real symmetric matrices, whereas $Q$ is antisymmetric. We see that all the independent components of the Riemann tensor are contained in the (3$\times$3)-matrices $M$, $N$, $P$, $Q$ and the scalar $R$. This suggests to introduce a further representation of the curvature tensor by using only (3$\times$3)-matrices. Indeed, noting that
(\ref{so31}) represents the irreducible pieces of the curvature with respect to the Lorentz group $SO(3,1)$ that is isomorphic to the group $SO(3,C)$, it is possible to introduce a local complex basis where the curvature is given as a (3$\times$3)-matrix. This is the so-called $SO(3,C)$-representation of the Riemann tensor whose irreducible pieces can be expressed as \cite{deb56,quev92}

\begin{subequations}
\begin{align}
{\cal R}&=W+E+S\,,\\
W&=M+iN\,,\\
E&=P+iQ\,, \\
S&=\frac{1}{12} R\,I_3\,.
\end{align}
\end{subequations}

Notice that the Einstein equations can be written as algebraic equations in this representation. For instance, in the case of vacuum spacetime we have that $E=0$ and $S=0$ and the vanishing of the Ricci tensor in terms of the components of the Riemann tensor corresponds to the algebraic condition ${\rm Tr}(W)=0$. The matrix $W$ has therefore only ten independent components, the matrix $E$ is hermitian with nine independent components and the scalar piece $R$ has only one component.

The eigenvalues of the curvature tensor correspond to the eigenvalues of the matrix ${\cal R}$. In general, they are complex $\lambda_n=a_n+ib_n$ with $n=1,2,3$. It is with respect to these curvature eigenvalues that the Petrov classification of gravitational fields is carried out.
As mentioned above, in the case of a vacuum solution the curvature matrix is traceless and hence the eigenvalues must satisfy the condition $\lambda_1+\lambda_2+\lambda_3=0$. Moreover, the case $\lambda_1\neq \lambda_2$ is the most general type in Petrov's classification and is called type I. Vacuum solutions with  $\lambda_1=\lambda_2$  correspond to type D gravitational fields. All the naked singularities we will study here belong to type I or D \cite{solutions}.

Since curvature eigenvalues characterize in an invariant manner the gravitational field in Petrov's classification, we propose to use them to identify the repulsive behavior of gravity. Our motivation is based upon the intuitive idea that a change in the gravitational field must generate a change at the level of the eigenvalues, since curvature is a measure of the gravitational interaction. If repulsive gravity is interpreted as the ``opposite" of attractive gravity, the curvature eigenvalues should be able to reproduce this behavior. In the case of compact gravitational sources, e.g.  black holes or naked singularities, the idea is to assume that the eigenvalues vanish at spatial infinity as a consequence of the asymptotic flatness condition. As we approach the central source, the eigenvalues will increase in value until they become infinity at the singularity. If gravity is everywhere attractive, one would expect that the eigenvalues increase monotonically from zero at spatial infinity until they reach their maximum value near the curvature singularity. However, if gravity becomes repulsive in some region, one would expect a different behavior for the eigenvalues; in particular, if repulsive gravity becomes dominant at a given point, one would expect at that point a change in the sign of at least one eigenvalue. The point where the eigenvalue vanishes would indicate that attractive gravity becomes entirely compensated by the action of repulsive gravity. 
This implies that the eigenvalue must have an extremal at some point before it changes its sign. According to this intuitive analysis, we propose to define the radius of repulsion $r_{rep}$ as the first extremal that appears in a curvature eigenvalue as we approach the origin of coordinates from infinity, i.e.,
\be
\frac{\partial \lambda}{\partial r}\bigg| _{r=r_{rep}}=0\ ,
\label{rep}
\ee
where $\lambda$ is any curvature eigenvalue and $r$ is a radial coordinate. In other words, the repulsion radius is the radial distance from the origin to the position of farthest extremal.  This radius determines the region where it should be possible to detect the effects of repulsive gravity, for instance, by using test particles. 

We will see in the following sections that the above purely intuitive motivation, which is the basis of our definition of repulsion radius, leads to physically reasonable results in several examples.

%%%%%%%%%%%%%%%%%%%%%%%%%%%%%%%%%%%%%%%%%%%%%%%%%%%%%%%%%%%%%%%%
%%%%%%%%%%%%%%%%%%%%%%%%%%%%%%%%%%%%%%%%%%%%%%%%%%%%%%%%%%%%%%%%
\section{Naked singularities with black hole counterparts}
\label{sec:bhs}

To study the structure of the curvature tensor of the Kerr-Newman naked singularities represented by the line element (\ref{kn}), it is convenient to introduce the orthonormal tetrad
\begin{subequations}
\begin{align}
\vartheta^0 = &   \left(\frac{r^2 - 2Mr + a^2+ Q^2} {r^2 + a^2\cos^2 \theta}\right)^{1/2} (dt - a\sin^2\theta d\varphi) \ ,   \nonumber\\
\vartheta^1 = &   {\frac{\sin\theta} {(r^2 + a^2\cos^2 \theta)^{1/2}}}[(r^2 + a^2) d\varphi - adt] \ , \nonumber\\
\vartheta^2 = &   (r^2 + a^2\cos^2 \theta)^{1/2} d\theta\ , \nonumber\\
\vartheta^3 = &    \left(\frac {r^2 + a^2\cos^2 \theta}{r^2 - 2Mr + a^2+ Q^2}\right)^{1/2} dr    \ .
\label{tetrad}
\end{align}
\end{subequations}

Thus, using the Cartan equations and the matrix formalism presented in the previous section, is is straightforward to show that the $(3\times 3)$ curvature matrix can be written as
\be
{\cal R}=\left(
         \begin{array}{ccc}
           l & \ 0 & 0 \\
           0 & \ l & 0 \\
           0 & \ 0 & -2 l + k \\
         \end{array}
       \right),
\label{knc}
\ee
with
\be
l= - \left[ M-{\frac {{Q}^{2} \left( r+ia\cos  \theta
 \right) }{ {r}^{2}+{a}^{2}  \cos^2  \theta
^{2}}} \right]  \left( \frac { r -ia\cos  \theta   }{
  {r}^{2}+{a}^{2} \cos^2 \theta } \right) ^{3} \ ,
\label{funl}
\ee
\be
k = -{\frac {{Q}^{2}}{ \left( {r}^{2}+{a}^{2}  \cos^2  \theta
    \right) ^{2}}} \ .
\label{funk}
\ee
Notice that in this invariant representation of the curvature all the important properties of the spacetime can easily be seen: It is asymptotically flat because $\lim_{r\to\infty} l = 0 = \lim_{r\to\infty} k$; it is flat in the limit $M=Q=a=0$; the only singular surface is defined by the equation $r^2 + a^2\cos^2\theta =0$; and the curvature does not suffer any particular change on the horizon.

%%%%%%%%%%%%%%%%%%  Dacha novoe

Consider the simplest case of the Schwarzschild spacetime. The $SO(3,C)$ curvature matrix reduces to
\be
{\cal R}_{Schw} = -\frac{M}{r^3} {\rm diag}(1,1,-2)\ ,
\ee
and so  the eigenvalues are
\be
\lambda_3 = \frac{2M}{r^3}\ ,\quad \lambda_1 = \lambda_2 = - \frac{M}{r^3}=-\frac{1}{2}\lambda_3\,.
\ee
All these eigenvalues as well as the corresponding first derivatives are monotonically increasing, do not change their sign, and diverge at the central singularity situated at $r=0$.  This means that the field is always attractive if we assume a positive mass. The only way to change the sign of these eigenvalues is to change the sign of the mass, $M\rightarrow -M$, leading to the vanishing of the black hole horizon. This implies that the spacetime with negative mass represents a naked singularity whose gravitational field is always repulsive. This is in accordance with our standard interpretation of the Schwarzschild spacetime.

\begin{figure}
\includegraphics[scale=1]{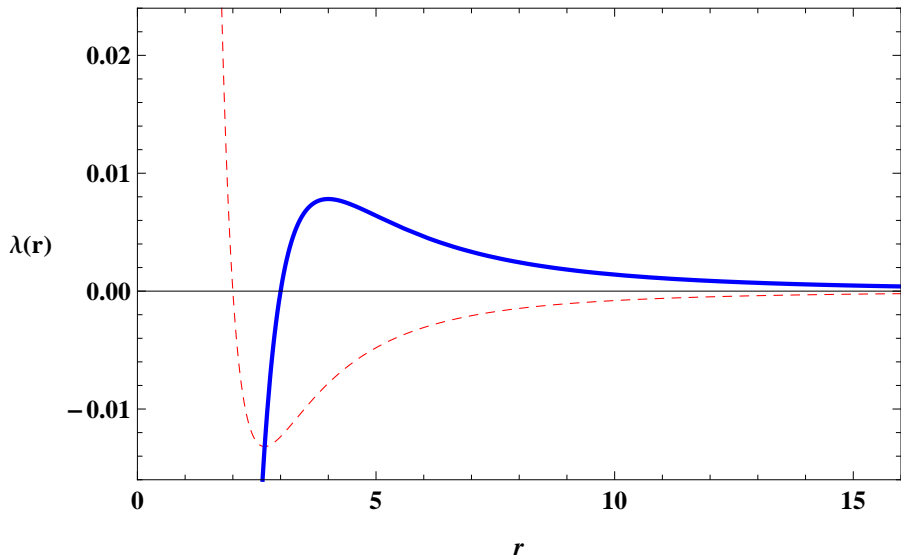}
\caption{Behavior of the curvature eigenvalues $\lambda_1$ (dashed curve) and $\lambda_3$ (solid curve) in terms of the radial coordinate $r$ for the values $M=1$ and $Q^2=2$ that correspond to a naked singularity}
\label{fig1}
\end{figure}

Consider now the Reissner-Nordstr\"om case. The curvature matrix can be represented as
\be
{\cal R}_{RN} = -\frac{1}{r^3}\left(M-\frac{Q^2}{r}\right){\rm diag}(1,1,-2)+\frac{Q^2}{r^4}{\rm diag}(0,0,-1)\ ,
\label{curvrn}
\ee
and the eigenvalues are
\be
\lambda_1=\lambda_2 = -\frac{M}{r^3} + \frac{Q^2}{r^4} \ , \quad \lambda_3 = \frac{2M}{r^3} - \frac{3Q^2}{r^4}=-2\lambda_1-\frac{Q^2}{r^4}\,.
\ee
The behavior of these eigenvalues is depicted in Fig. \ref{fig1}.
Computing their first derivatives,  it is easy to see that $\lambda_1$ and $\lambda_3$ have extremals at $\frac{4Q^2}{3M}$ and at $\frac{2Q^2}{M}$ respectively. Thus, according to  our  definition of repulsion radius (\ref{rep}),  we take the greatest value
\be
r^{^{RN}}_{rep} =\frac{2Q^2}{M} \ ,
\ee
as defining the repulsion radius of the Reissner-Nordstr\"om spacetime.

In contrast with the value obtained from the Newtonian limit in Sec. \ref{sec:meff}, this value of the repulsion radius is greater than the classical radius of a charged body with mass $M$ and charge $Q$. It is therefore possible to interpret repulsive gravity in this case as a classical effect.

\begin{figure}
\includegraphics[scale=1]{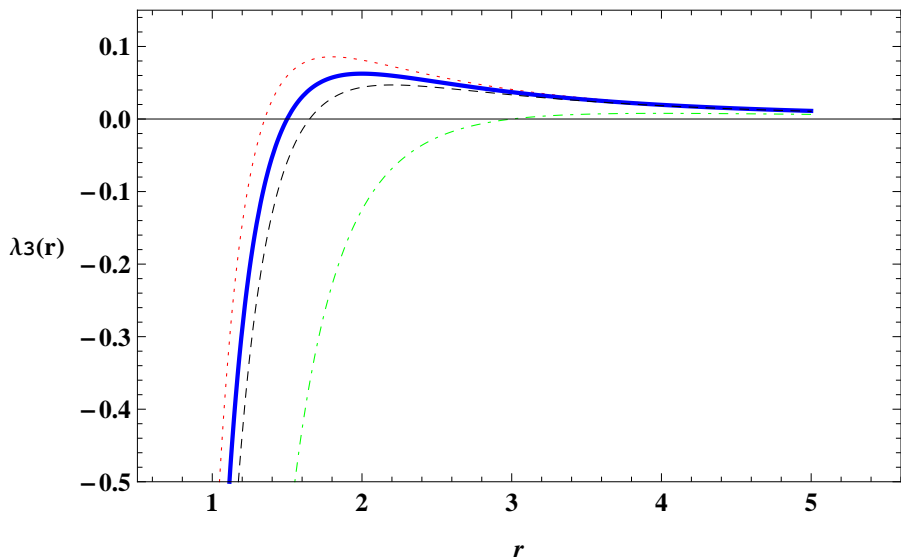}
\caption{Behavior of the curvature eigenvalues $\lambda_3$ in terms of the radial coordinate $r$ for $M=1$ and $Q=\sqrt{0.9}$ (red dotted curve), $Q=1$ (blue solid curve), $Q=\sqrt{1.1}$ (black dashed curve) and $Q=\sqrt{1.2}$ (green dash-dotted curve).}
\label{fig2}
\end{figure}

In Fig. \ref{fig2}, we show the behavior of the eigenvalue $\lambda_3$, which determines the repulsion radius, for a fixed value of the mass and different values of the charge.
We notice that even in the case of black holes a repulsion radius exists. This is in accordance with the repulsive effects that have been detected by analyzing the motion of test particles  in the Reissner-Nordstr\"om black hole \cite{repRN,pqr11a}.
In fact, the outer horizon radius can be expressed as
\be
r_+ = M\left( 1 + \sqrt{1-\frac{r_{rep}^{^{RN}}}{2M}} \right)
\ee
in terms of the repulsion radius. We see that a black hole can exist only if $r^{^{RN}}_{rep}\leq 2M$, and naked singularities are characterized by $r^{^{RN}}_{rep}> 2M$. The two radii coincide at $r_+ = r_{rep}^{^{RN}} = \frac{3}{2}M$, i.e. for $\frac{Q^2}{M^2} = \frac{3}{4}$. Accordingly, for values of the charge-to-mass ratio $\frac{Q^2}{M^2}\leq \frac{3}{4}$, the repulsion radius is completely covered by the horizon. For values in the interval $1\geq\frac{Q^2}{M^2}>\frac{3}{4}$, the repulsion radius is located outside the horizon. Finally, for  naked singularities with $1<\frac{Q^2}{M^2}$ only the  repulsion radius exists. This is in agreement with the results obtained by analyzing the circular motion of test particles. In fact, in \cite{pqr11a} it was found that due to repulsive gravity a test particle situated on the radius $\frac{Q^2}{M}$
can stay at rest with respect to an observer at infinity. The place where the test particle can stay at rest is located inside the sphere defined by the repulsion radius.

\begin{figure}
\includegraphics[scale=0.3]{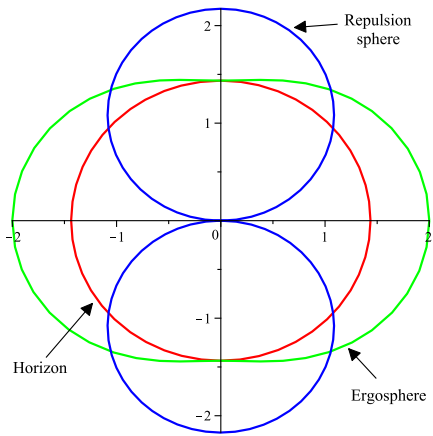}
\caption{Region determined by the repulsion radius $r_{rep}^{^K}=\left(1+\sqrt{2}\right)a\cos\theta$ in a Kerr spacetime with $a=1$.
Only the projection of the repulsion spheres on a plane orthogonal to the equatorial plane is here illustrated.  }
\label{fig3}
\end{figure}

Consider now the case of the Kerr spacetime. The curvature matrix is given by
\be
{\cal R}= -M\left(\frac{r- i a \cos\theta}{r^2+a^2\cos^2\theta}\right)^3{\rm diag}(1,1,-2) \ ,
\label{kerr}
\ee
with eigenvalues
\be
\lambda_1=\lambda_2 = -\frac{1}{2}\lambda_3 = -M\left(\frac{r- i a \cos\theta}{r^2+a^2\cos^2\theta}\right)^3 \ .
\ee
Computing the partial derivatives of the real and imaginary part of the eigenvalues, one can see that there are several extremal points. We choose the greatest value to define the repulsion radius
\be
r_{rep}^{^K} = \left(1+\sqrt{2}\right) a \cos\theta\approx2.41a \cos\theta\,,
\ee
of the Kerr central source. The configuration determined by this repulsion radius corresponds to two spheres whose centers are located on the symmetry axis at a distance $\frac{1}{2}(1+\sqrt{2}) a $ from the equatorial plane (see Fig. \ref{fig3}).

Notice that on the equatorial plane ($\cos\theta=0$)  the repulsion radius vanishes. This is in accordance with the behavior of the curvature eigenvalues (cf. Eq.(\ref{kerr})) which on the equatorial plane coincide with those of the Schwarzschild spacetime. This, however, does not mean that on the equatorial plane repulsive gravity is absent. In fact, we interpret the repulsion spheres as the regions where repulsive gravity is ``generated'', and can become dominant. Of course, outside these regions, the effects of repulsive gravity can also be detected \cite{repkerr1,repkerr2}. This has been recently confirmed by studying the circular motion of test particles on the equatorial plane of the Kerr spacetime \cite{pqr11b}.

Notice also that the Kerr repulsion radius vanishes as the angular momentum vanishes, independently of the value of the mass. This result is in agreement with the result obtained in the Schwarzschild case. In fact, we have seen that a positive mass cannot generate repulsive gravity and therefore the limiting case of a  vanishing angular momentum with positive mass cannot correspond to a repulsion radius. This means that a positive mass alone cannot generate repulsive gravity, but only in connection with an angular momentum it can act as a source of repulsive gravity.

Using the expression for the repulsion radius, it is possible to rewrite the radius of the outer ergosphere as
\be
\frac{r_{erg}}{M}= 1 + \left[1-\left(\frac{r_{rep}^{^K}}{(1+\sqrt{2})M}\right)^2\right]^{1/2} \ ,
\ee
so that the two radii coincide at
\be
r_{rep}^{^K} = r_{erg} = \frac{(1+\sqrt{2})^2}{2+\sqrt{2}}M\approx 1.71 M\,.
\ee
This implies that the repulsion spheres can be entirely contained inside the ergosphere, depending on the value of the angular momentum. Nevertheless, there exists a wide interval of values for black holes and naked singularities in which parts of the repulsion spheres can be situated outside the ergosphere.

\begin{figure}
\includegraphics[scale=1]{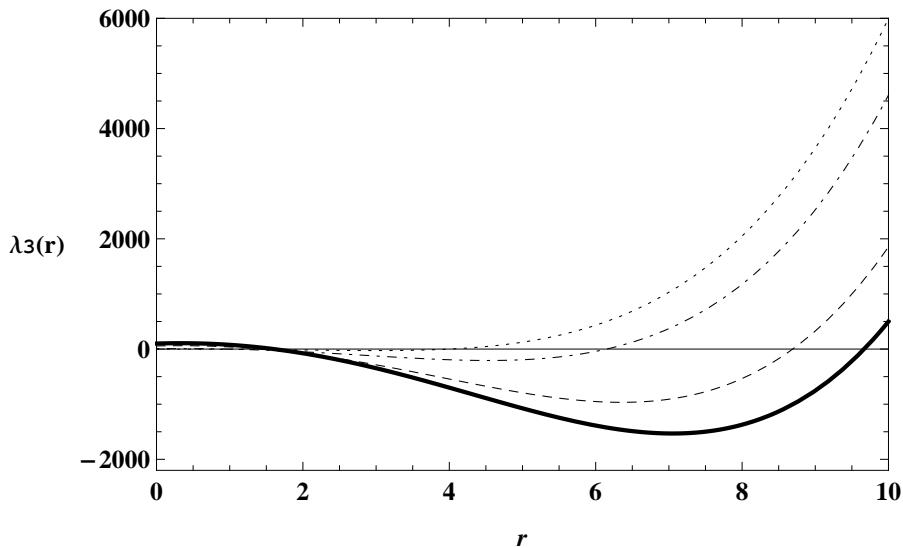}
\caption{The zeros of the polynomial (\ref{kncond}) determine the repulsion radius of a Kerr-Newman naked singularity with $M=1$, $Q=\sqrt{2}$, and $a=\sqrt{10}$, for different values of the azimuthal angle:
$\theta=0$ (solid curve), $\theta=\frac{\pi}{6}$ (long-dashed curve), $\theta=\frac{\pi}{3}$ (dash-dotted curve), $\theta=\frac{\pi}{2}$ (dotted curve).  }
\label{fig4}
\end{figure}

We finally analyze the case of the Kerr-Newman spacetime. The curvature matrix is given in Eq.(\ref{knc}), and the eigenvalues are
\be
\lambda_1 = \lambda_2 = l\ , \quad \lambda_3 = -2l + k=-2\lambda_1+k \,,
\ee
with $l$ and $k$ given in Eqs.(\ref{funl}) and (\ref{funk}), respectively. A numerical analysis shows that all the eigenvalues have extremals, and the one of the real part of $\lambda_3$ is the first found when approaching the origin of coordinates from infinity. The real part of $\lambda_3$ has extremals at the roots of the equation
\be
r^3(Mr - 2Q^2) + a^2\cos^2\theta [2Q^2 r + M(a^2 \cos^2\theta - 6 r^2)] = 0 \ .
\label{kncond}
\ee
In the limiting case of a vanishing angular momentum $(a=0)$, we obtain  the Reissner-Nordstr\"om repulsion radius,
$r_{rep}^{^{RN}} = \frac{2 Q^2}{M}$, and for a vanishing electric charge we recover the expression for the Kerr repulsion radius $r_{rep}^{^K} = (1+\sqrt{2}) a \cos\theta$. In general, the roots will depend on the explicit values of $Q$ and $a$. In Fig. \ref{fig4}, we find numerically the zeros of the polynomial (\ref{kncond}) for a particular naked singularity, indicating that the repulsion radius exists.

\begin{figure}
\includegraphics[scale=0.4]{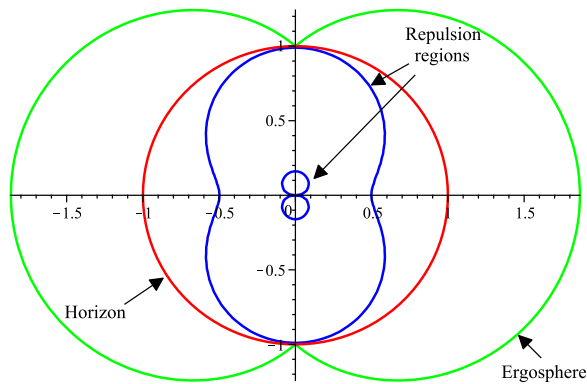}
\caption{Location of the repulsion regions of a Kerr-Newman extreme black hole with $M=1$, $Q=\frac{1}{2}$ and $a=\sqrt{\frac{3}{4}}$. The repulsion region closer to the central singularity exists only for $\theta\neq 0$, and corresponds to a second root of the polynomial (\ref{kncond}) which exists only outside the equator. }
\label{fig5}
\end{figure}

We see that on the equator, the repulsion radius coincides with the Reissner-Nordstr\"om radius. Then, it increases as the azimuthal angle decreases and reaches its maximum value on the axis, $\theta=0$. A more detailed numerical analysis shows the presence of a second repulsion radius at distances very close to the singularity. We illustrate this behavior in Fig. \ref{fig5} for a particular extreme black hole. It follows that in this case each repulsion region is represented by two intersecting spheroids whose radius is proportional to the value of  the angular momentum parameter $a$. 
The intersection of the spheroids with the equatorial plane determines a circle whose radius coincides with  the Reissner-Nordstr\"om repulsion radius, $r_{rep}^{^{RN}}=\frac{2Q^2}{M}$. This means that as the value of the electric charge tends to zero, the intersecting circle must vanish, and the repulsion regions  turn into two spheres with only one point of intersection on the origin of coordinates. This coincides with the result obtained for the Kerr spacetime, as illustrated in Fig. \ref{fig3}. In the case of naked singularities, the geometric structure of the repulsion regions remains unchanged for small values of $Q$. In the limit $Q\rightarrow \infty$, however, the outer repulsion regions turn into a sphere, whereas the interior repulsion region still corresponds to two spheroids with no intersection on the equatorial plane.

Our numerical study of the Kerr-Newman repulsion radius leads to results that are in agreement with the limiting cases in which the results can be obtained analytically. In addition, the idea of determining regions where repulsive effects occur may represent a criterion to investigate how to match interior and
exterior solutions of Einstein's equations \cite{quev12}.

%%%%%%%%%%%%%%%%%%%%%%%%%%%%%%%%%%%%%%%%%%%%%%%%%%%%%%%%%%%%%%%%
%%%%%%%%%%%%%%%%%%%%%%%%%%%%%%%%%%%%%%%%%%%%%%%%%%%%%%%%%%%%%%%%
\section{Final outlooks and perspectives}
\label{sec:con}

In this work, we proposed an invariant definition of the repulsion radius in terms of the eigenvalues of the curvature tensor. It is defined as the radial distance between the origin of coordinates and the first extremal that is found in any curvature eigenvalue when approaching the origin of coordinates from infinity. The space contained within the repulsion radius can be interpreted as the region where the effects due to repulsive gravity can become very important.

We tested our invariant definition of the repulsive gravity in all the naked singularity spacetimes with a black hole counterpart. In the case of the Schwarzschild metric we established that repulsive gravity can only exist for negative values of the mass parameter, a result  which is in agreement with our intuitive representation of a Schwarzschild naked singularity. In the case of a Reissner-Nordstr\"om spacetime we found an analytic expression for the repulsion radius, $r_{rep}^{^{RN}}=\frac{2Q^2}{M}$. In this case, for black holes with $\frac{Q^2}{M^2}\leq \frac{3}{4}$ the repulsion sphere is covered by the horizon. Otherwise, the repulsion sphere is located outside the horizon.

We derived for the Kerr spacetime the repulsion radius $r_{rep}^{^K} = (1+\sqrt{2})a \cos\theta$, determining  a configuration of two spheres that can be located either completely or partially inside the ergosphere. In the case of the Kerr-Newman spacetime, it was not possible to derive an analytic expression for the repulsion radius, but we performed a detailed numerical analysis of the corresponding conditions. We found that in this case the repulsion region consists of two intersecting ellipsoids that generate a geometric structure which is symmetric with respect to the equatorial plane and to the azimuthal axis. This structure reduces to the corresponding configurations in the limiting Kerr and Reissner-Nordstr\"om cases.

We notice that the invariant definition of repulsion radius as presented in this work is not unique. One could, for instance, define an alternative radius as the largest distance from zero of the radial integral of the curvature eigenvalues which vanish at infinity\footnote{We thank an anonymous referee for pointing out this possibility.}. A straightforward computation shows that the integral of the real part of $\lambda_3$ yields the largest alternative repulsion radius which can be expressed as
\be
\tilde r _{rep}^{^{KN}} = 
{\frac {{Q}^{2}+\sqrt {{Q}^{4}+4\,{a}^{2} M^2 \cos^2\theta 
}}
{2M}}
\ee 
for the Kerr-Newman spacetime. The technical advantage of this alternative definition is that it leads to an analytic expression also in the 
Kerr-Newman spacetime, instead of the numerical approach that must be used when applying our definition as given in Eq.(\ref{rep}). 
However, there is a particular difference between the two radii, suggesting that our definition could lead to more physically reasonable results. In fact, we can see that $\tilde r_{rep}$ as well as  $r_{rep}$ define a similar geometric structure for the repulsion regions, but in general the only difference is that $\tilde r_{rep}< r_{rep}$ in all the cases (the exterior repulsion radius in the Kerr-Newman spacetime). So, for instance, in
 the Reissner-Nordstr\"om spacetime, the alternative repulsion radius  coincides with the classical radius $Q^2/M$. This choice, however, does not agree with the results obtained by analyzing the motion of test particles in this spacetime. In fact, in \cite{pqr11a} it was shown that a particle located at the classical radius can remain ``at rest" (zero angular momentum) with respect to an observer situated at infinity. This means that the classical radius corresponds to the place where repulsive gravity entirely compensates attractive gravity and, therefore, cannot be considered as a definition for the onset of repulsion. Moreover, inside the classical radius no timelike circular orbits are allowed. It is also interesting to note that at the classical radius the Weyl part of the curvature tensor 
(\ref{curvrn}) vanishes, which could be interpreted as a spot of ``zero gravity". The definition proposed here in Eq.(\ref{rep}) yields in this case 
$r_{rep}^{^{RN}} = 2Q^2/M$, which is greater than the classical radius and, therefore,  more suitable to be used as the definition of the onset of repulsive gravity.

The repulsion region as defined here is located in all the cases near the origin of coordinates so that repulsive effects are expected to become very important only in the vicinity of the central compact object. This result is confirmed by a different study in which the circular motion of test particles has been investigated for black holes and naked singularities \cite{pqr11a,pqr11b,pqr13}.

%%%%%%%%%%%%%%%%%%%%%%%%%%%%%%%%%%%%%%%%%%%%%%%%%%%%%%%%%%%%%%%%%%%%%%
\section*{Acknowledgements}

This work was  supported by DGAPA-UNAM, Grant No. 113514, and Conacyt-Mexico, Grant No. 166391. O.L. is financially supported by the European PONa3 00038F1 KM3NET (INFN) Project.

%%%%%%%%%%%%%%%%%%%%%%%%%%%%%%%%%%%%%%%%%%%%%%%%%%%%%%%%%%%%%%%%%%%%%%%%%%%%%%%%%%%%%%%

\end{document}